\documentclass[12pt,a4paper]{article}

\usepackage[latin1]{inputenc}
\usepackage[T1]{fontenc}
\usepackage[english,american]{babel}
\usepackage{lmodern}

\usepackage[DIV=12]{typearea}

\usepackage[reqno]{amsmath} %gives new environments such as split, align etc. 
\usepackage{bbm} %for sets etc. $\mathbbm{N}$

\usepackage{array}
\usepackage{multirow}

\usepackage{graphicx}
\usepackage{float}
\usepackage{rotating}
\usepackage{caption}
\usepackage{subfig}
\usepackage{color}

\usepackage[numbers,sort&compress]{natbib}

\usepackage{hyperref} 

\definecolor{darkblue}{rgb}{0,0,.5}
\hypersetup{colorlinks=true, breaklinks=true, linkcolor=darkblue, citecolor=darkblue, filecolor=darkblue, menucolor=darkblue, pagecolor=darkblue, urlcolor=darkblue}

\usepackage{hypernat}

\usepackage[all]{hypcap} 
\def\ra{\rightarrow} 
\def\be{\begin{equation}} 
\def\ee{\end{equation}}
\def\bea{\begin{eqnarray}} 
\def\eea{\end{eqnarray}}

\newcommand{\obb}{0\mbox{$\nu\beta\beta$ decay}} 
\newcommand{\zbb}{2\mbox{$\nu\beta\beta$ decay}}

\newcommand{\nel}{\mbox{$\nu_e$} }

\newcommand{\ema}{\mbox{$\langle m_{\nu_e} \rangle $}}

\begin{document}
%%%%%%%%%%%%%%%%%%%%%%%%%%%%%%%%%%%%%%%%%%%%%%%%%%%%%%%%%%%%%%%%%%%%%%%%%%%%%%%
%%%%%%%%%%%%%%%%%%%%%%%%%%%%%%%%%%%%%%%%%%%%%%%%%%%%%%%%%%%%%%%%%%%%%%%%%%%%%%%

\begin{titlepage}
\title{\vspace*{-2.0cm}
%\hfill {\small \today}\\[20mm]
\bf\Large
Consistency test of neutrinoless double beta decay with one isotope\\[5mm]\ }

\author{
Michael Duerr$^a$\thanks{email: \tt michael.duerr@mpi-hd.mpg.de}~,~Manfred Lindner$^a$\thanks{email: \tt manfred.lindner@mpi-hd.mpg.de}~,~and
Kai Zuber$^{b}$\thanks{email: \tt zuber@physik.tu-dresden.de}
\\ \\
$^a${\normalsize \it Max-Planck-Institut f\"ur Kernphysik,}\\
{\normalsize \it Postfach 10 39 80, 69029 Heidelberg, Germany}\\
\\
$^b${\normalsize \it Technical University Dresden, Institut f\"ur Kern- und Teilchenphysik,}\\
{\normalsize \it 01069 Dresden, Germany}\\
}
\date{}
\maketitle

% no page numbering or other automatic things on the first page
\thispagestyle{empty}

\begin{abstract}
\noindent
We discuss a consistency test which makes it possible to discriminate unknown nuclear background lines from neutrinoless double beta decay with only one isotope. By considering both the transition to the ground state and to the first excited $0^+$ state, a sufficiently large detector can reveal if neutrinoless double beta decay or some other nuclear physics process is at work. Such a detector could therefore simultaneously provide a consistency test for a certain range of Majorana masses and be sensitive to lower values of the effective Majorana mass \ema.

\vspace{1em}


\vspace{1em}

\end{abstract}

\end{titlepage}
\section{Introduction} 
\noindent  
The first solid evidence for New Physics beyond the Standard Model is the proof that neutrinos have a nonvanishing rest mass, as could be convincingly shown in various neutrino oscillation experiments \cite{GonzalezGarcia:2007ib,Mohapatra:2005wg}. The results show the violation of individual flavor lepton number, while conserving total lepton number. A violation of the latter would have even deeper consequences for our understanding of the Universe. The gold-plated process demonstrating total lepton number violation is neutrinoless double beta decay of atomic nuclei. 

Double beta decay is characterized by a change in atomic number $Z$ by two units, while leaving the mass number $A$ constant. It is observable for 35 even-even nuclei, as single beta decay is energetically forbidden or at least strongly suppressed. It may occur in the two-neutrino mode as well as in the neutrinoless mode:
\begin{equation}
(Z,A) \rightarrow (Z+2,A) + 2 e^- + 2 \bar{\nu}_e \quad (\zbb)
\end{equation}
and
\begin{equation}\label{eq:proc0nu}
(Z,A) \rightarrow (Z+2,A) + 2 e^-  \quad (\obb) \, .
\end{equation}

\zbb\ conserves lepton number and is therefore allowed in the Standard Model. However, this transition is second order in the weak Hamiltonian and thus strongly suppressed. Nevertheless, it has been detected experimentally for about a dozen isotopes, and their half-lives have been measured. Depending on the nuclide under consideration, half-lives are of the order $10^{20}\,\mathrm{y}$. 

The second process, \obb, is forbidden in the Standard Model
as it violates total lepton number by two units. It can be seen as two subsequent steps (''Racah sequence''~\cite{Racah:1937}):
\begin{equation}
 \begin{split}
 (Z,A) &\rightarrow (Z+1,A) + e^- +  \bar{\nu}_e \\
 (Z+1,A) + \nel &\rightarrow (Z+2,A) + e^- \, ,
 \end{split}
\end{equation}
which is only possible if the neutrino is its own antiparticle (a so-called Majorana particle) and if it has a nonvanishing rest mass to account for the
helicity matching. As the decay is a kind of a black box, any $\Delta L = 2$ process could contribute to \obb . \obb\ has not been observed experimentally yet.\footnote{Note that one claim for a positive signal of $0\nu\beta\beta$ decay exists. A subgroup of the Heidelberg--Moscow collaboration gives the half-life $T^{0\nu}_{1/2} = 1.98\times 10^{25}\,\mathrm{y}$, where the $3\sigma$ range is given by $(1.04$--$20.38)\times 10^{25}\,\mathrm{y}$~\cite{KlapdorKleingrothaus:2004wj}, and a more recent analysis 
results in  $T^{0\nu}_{1/2} = \left(2.23^{+0.44}_{-0.31}\right)\times 10^{25}\,\mathrm{y}$~\cite{KlapdorKleingrothaus:2006ff}.} Nevertheless, stringent bounds on the half-lives of various elements have been extracted from experiments. Best limits are of the order $10^{25}\,\mathrm{y}$.

The total decay rate of neutrinoless double beta decay (if mediated by light neutrino exchange) is given by~\cite{Bilenky:2010kd}
\begin{equation}\label{eq:decay_rate}
 \frac{\Gamma^{0\nu}}{\ln 2} = \frac{1}{T^{0\nu}_{1/2}} = \ema^2 \left|\mathcal{M}^{0\nu}\right|^2 G^{0\nu}(Q,Z)\, .
\end{equation}
Here, $T^{0\nu}_{1/2}$ is the half-life of $0\nu\beta\beta$. The nuclear matrix element $\mathcal{M}^{0\nu}$ and the phase space integral $G^{0\nu}(Q,Z)$ depend on the nucleus under consideration. In \obb, it is thus possible to measure \ema, the so-called effective Majorana mass of the electron neutrino. For light neutrinos, it is given by 
\begin{equation}\label{eq:ema}
\ema = \left| \sum_i U_{ei}^2 m_i \right| =  \left| \sum_i \left| U_{ei} \right|^2 e^{2i\alpha_i} m_i \right|\, ,
\end{equation}
where $U_{ei}$ are the elements of the first line of the Pontecorvo--Maki--Nakagawa--Sakata neutrino mixing matrix, and $\alpha_i$ are the two Majorana $CP$-violating phases.
In case of $CP$-invariance ($\alpha_i = 0, \pi$), we may write
\begin{equation}
\ema = \left| m_1 U_{e1}^2 \pm m_2  U_{e2}^2  \pm m_3  U_{e3}^2 \right|\,  .
\end{equation}

\begin{figure}
 \centering
\includegraphics[width=0.7\linewidth]{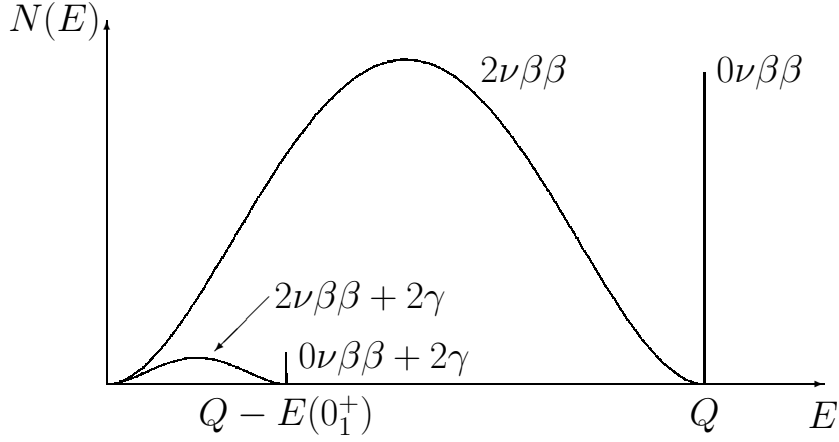}
\caption{Schematic plot of the sum energy spectrum of the emitted electrons for the various decays to be discussed in this paper. As emitted neutrinos may take away an arbitrary amount of energy, the spectrum for $2\nu\beta\beta$ decay is continuous, whereas the spectrum of $0\nu\beta\beta$ decay is a single line at the maximum energy $Q$. The ratio between both is not to scale. For the decay to an excited final state, the diagrams are qualitatively the same. However, as energy is taken away by the emitted photons, the line for $0\nu\beta\beta$ decay lies at lower energies. Moreover, the number of decays to excited states is lower than the number of decays to the ground state.}\label{fig:energy_spectrum}
\end{figure}

The experimental signal of \obb\ is two electrons in the final state, whose
energies add up to the $Q$~value of the nuclear transition. For \zbb, however,
the sum energy spectrum of both electrons will be continuous as the neutrinos may take away an arbitrary amount of energy (see Fig.~\ref{fig:energy_spectrum}). The total decay rates, and hence the inverse half-lives, are a strong function of the available $Q$~value. The rate of \obb\ scales with $Q^5$ compared to a $Q^{11}$ dependence for \zbb, due to the dependence on the phase space factor $G^{0\nu}(Q,Z)$ [cf.\ Eq.~(\ref{eq:decay_rate})]. Therefore, isotopes with a high $Q$~value (above about 2 MeV) are usually considered for experiments on \obb. This restricts the candidates to 11 promising isotopes, which are given in Table~\ref{tab:exstate-alliso} together with the corresponding $Q$~values. For recent reviews on double beta decay, see~\cite{Zuber:2006hv,Elliott:2004hr,Avignone:2007fu,Bilenky:2010kd}.

\renewcommand{\arraystretch}{1.5}
\begin{sidewaystable}
\centering
\caption{All double beta emitters with a $Q$~value larger than 2 MeV and their corresponding $Q$~values. All $Q$~values with an error larger than 1 keV are taken from~\cite{Audi:2002rp}; all others were recently remeasured using Penning traps \cite{Kolhinen:2010zz,Rahaman:2007ng,Redshaw:2009zz,Redshaw:2007un,McCowan:2010zz}. Furthermore, the energy of the first excited $0^+$ state as taken
from~\cite{toi:1998} is shown. All nuclear matrix elements are obtained within the IBM-2 model~\cite{Barea:2009zza,iac2010,iac2011}.
As can be seen from pure phase space considerations, $^{150}$Nd and $^{96}$Zr would be the choice; however, including the 
matrix elements, $^{150}$Nd and $^{76}$Ge seem to be the most promising cases to study.}\label{tab:exstate-alliso}

\begin{tabular}{cccccccc}
\hline
Decay mode & $Q \; [\mathrm{keV}]$ & $E (0_1^+) \; [\mathrm{keV}]$ & $\mathcal{M}_{0\nu}^{\mathrm{g.s.}}$& $\mathcal{M}_{0\nu}^{0_1^+}$ & $ (Q-E (0_1^+))^5 /Q^5$ & $\left(\mathcal{M}_{0\nu}^{0_1^+} / \mathcal{M}_{0\nu}^{\mathrm{g.s.}}\right)^2$ & $\Gamma_{0_1^+} / \Gamma_{\mathrm{g.s.}}$ \\
%& (keV) & (\%) & (c/keV/kg/hr) & ($10^{18}$ yrs)\\
 \hline \hline
$_{20}^{48}$Ca$\rightarrow _{22}^{48}$Ti & $4274 \pm 4$ \cite{Audi:2002rp} & 2997 & - & - & $2.38 \times 10^{-3}$ &  - & - \\
$_{32}^{76}$Ge$\rightarrow _{34}^{76}$Se & $2039.04 \pm 0.16$ \cite{Rahaman:2007ng} & 1122 & 5.465 & 2.479 & $1.84 \times 10^{-2}$ & 0.206 & $3.79 \times 10^{-3}$\\
$_{34}^{82}$Se$\rightarrow _{36}^{82}$Kr & $2995.5 \pm 1.9$  \cite{Audi:2002rp}& 1488 & 4.412 & 1.247 & $3.23 \times 10^{-2}$ & 0.080 & $2.58 \times 10^{-3}$\\
$_{40}^{96}$Zr$\rightarrow _{42}^{96}$Mo & $3347.7 \pm2.2$  \cite{Audi:2002rp}& 1148 & 2.530 & 0.044 & $1.22 \times 10^{-1}$ & $3.02 \times 10^{-4}$& $3.70 \times 10^{-5}$\\
$_{~42}^{100}$Mo$\rightarrow _{~44}^{100}$Ru & $3034.40 \pm 0.17$ \cite{Rahaman:2007ng} & 1130 & 3.732 & 0.419 & $9.74 \times 10^{-2}$ & $1.26 \times 10^{-2}$ & $1.23 \times 10^{-3}$\\
$_{46}^{110}$Pd$\rightarrow _{48}^{110}$Cd & $2004 \pm 11$ \cite{Audi:2002rp}& 1473 & 3.623 & 1.599&  $1.31 \times 10^{-3}$ & 0.195 & $2.54 \times 10^{-4}$\\
$_{~48}^{116}$Cd$\rightarrow _{~50}^{116}$Sn & $2809 \pm 4$ \cite{Audi:2002rp} & 1757 & 2.782 & 1.047 &  $7.37 \times 10^{-3}$ & 0.142 & $1.04 \times 10^{-3}$\\
$_{50}^{124}$Sn$\rightarrow _{52}^{124}$Te & $2287.8 \pm 1.5$ \cite{Audi:2002rp}& 1657 & 3.532 & 2.721 &  $1.59 \times 10^{-3}$& 0.594 &  $9.46 \times 10^{-4}$ \\
$_{~52}^{130}$Te$\rightarrow _{~54}^{130}$Xe & $2527.518 \pm 0.013$ \cite{Redshaw:2009zz} & 1794 & 4.059 & 3.090 & $2.06 \times 10^{-3}$& 0.580 & $1.19 \times 10^{-3}$\\
$_{~54}^{136}$Xe$\rightarrow _{~56}^{136}$Ba & $2457.83 \pm 0.37$ \cite{Redshaw:2007un,McCowan:2010zz}&  1579 & 3.352 & 1.837 & $5.84 \times 10^{-3}$& 0.300 & $1.76 \times 10^{-3}$\\
$_{~60}^{150}$Nd$\rightarrow _{~62}^{150}$Sm & $3371.38 \pm 0.20$ \cite{Kolhinen:2010zz} &  740 & 2.321 & 0.395 & $2.90 \times 10^{-1}$& $2.90 \times 10^{-2}$&  $8.39 \times 10^{-3}$\\
\hline 
\end{tabular}
\end{sidewaystable}

\section{Double beta decay to excited \texorpdfstring{$\boldsymbol{0_1^+}$}{01+} states}
\begin{figure}
 \centering
\includegraphics[width=0.7\linewidth]{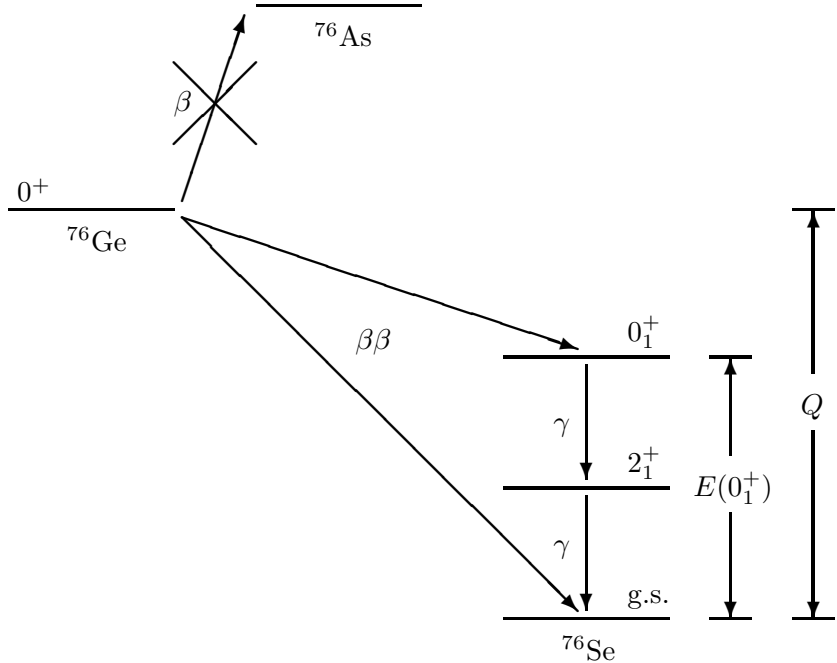}
\caption{Scheme of the double beta decay of $^{76}$Ge. Single beta decay to $^{76}$As is energetically forbidden, thus the only open decay channel is double beta decay. It may proceed in the two-neutrino mode, as well as in the neutrinoless mode. Additionally, we will discuss the decay to the first excited $0^+$ level.}\label{fig:level_scheme}
\end{figure}

Usually, double beta decay to the ground state (``g.s.'' in equations) of the final nucleus is considered. However, practically all interesting nuclei, i.e., those with 
a $Q$~value above 2 MeV (see Table~\ref{tab:exstate-alliso}) have at least one excited $0^+$ and one excited $2^+$ state which is accessible by double beta decay as
well. The level scheme of $^{76}$Ge is given in Fig.~\ref{fig:level_scheme} as an example. 
Transitions to excited $2^+$ states might be dominated by potential contributions of V+A interactions (see, however, \cite{Tomoda:1999zc}).

The decay rate to excited states is lower due to the lower $Q$~value of the decay. 
The ratio between the decay rate to the excited $0_1^+$ state and the ground state is given by
\begin{equation}
\frac{\Gamma_{0_1^+}}{\Gamma_{\mathrm{g.s.}}}= \frac{(Q-E(0_1^+))^n}{Q^n} \times \left(\frac{\mathcal{M}^{0_1^+}}{\mathcal{M}^{\mathrm{g.s.}}}\right)^2 \, ,
\end{equation}
where $E(0_1^+)$ is the energy of the first excited state with respect to the ground state.
For \obb, $n=5$ describes the phase space dependence, $\mathcal{M}_{0\nu}^{\mathrm{g.s.}}$ denotes the nuclear matrix element for the decay to the ground state, whereas $\mathcal{M}_{0\nu}^{0_1^+}$ denotes the nuclear matrix element for the decay to the first excited $0^+$ state. 
A similar relation holds for the \zbb\ mode, however with different matrix elements $\mathcal{M}_{2\nu}^{0_1^+}$ and $\mathcal{M}_{2\nu}^{\mathrm{g.s.}}$ 
as well as with a different scaling law for the phase space, i.e., $n=11$. 

So far, excited state transitions have only been observed in two nuclides, 
namely $^{100}$Mo \cite{Kidd:2009ai} and $^{150}$Nd \cite{Barabash:2009wy}, both considered to be \zbb.
The question arises whether, in case of an observation of the ground state transition within a single isotope, the transition to the first excited $0^+$ state
can be used as a consistency check, i.e., whether it is possible to prove \obb\ with one isotope in one experiment in two different ways. Such a test might
be desirable in future large-scale experiments due to the involved costs. The results for
all double beta emitters with a $Q$~value of at least 2 MeV are compiled in Table~\ref{tab:exstate-alliso}. 
For consistency, the matrix elements from a single method, the interacting boson model (IBM-2), have been used ~\cite{Barea:2009zza,iac2010,iac2011}. Thus, numbers might change slightly if other calculations are used. Unfortunately,
there is no complete set of such matrix elements available for all 11 isotopes, including also excited state transitions. 
It turns out that the two most suitable choices for such an internal consistency check would be  $^{150}$Nd and $^{76}$Ge, the first one about a 
factor of 2.21 better. 

Let us remark that a possible benefit of the proposed method is that the nuclear matrix elements
of the transition to the ground state and to the excited state may have
common uncertainties, which would cancel in the ratio. Improvement of the nuclear matrix elements will allow for a more precise extraction of the effective neutrino mass from half-life measurements.

\section{Experimental considerations}
The expected signature for the required second decay mode into the excited $0_1^+$ state will be two electrons and two gammas with defined 
energies in contrast to the ground state transition having only two electrons with a defined sum energy. Thus, the gammas
must be clearly separated from the emitted electrons in the experiment, otherwise they would look like a ground state transition. Hence, a purely calorimetric 
approach without spatial resolution to determine the individual gammas will fail. Consequently, in a large homogeneous detector, 
there must be spatial resolution to see the gammas independently from the emitted electrons. In a high granularity detector, 
the granularity should be chosen such that both gammas
might leave the crystal containing the decay without any interaction, making it possible to search for coincidences with high efficiency.

\begin{figure}
 \centering 
\includegraphics[width=0.6\textwidth]{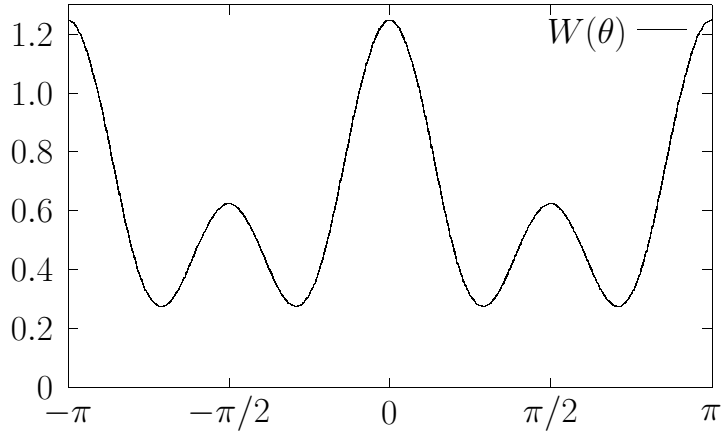}
\caption{Angular correlation from Eq.~(\ref{eq:angular_correlation}).}\label{fig:angular_correlation}
\end{figure}

As all double beta decays into the first excited $0^+$ state will de-excite via the sequence of $0^+ \ra 2^+ \ra 0^+$, there will also be 
a $\gamma \gamma$ angular correlation, which for the given angular momentum sequence is
\begin{equation}\label{eq:angular_correlation}
W (\theta) = \frac{5}{8} \times  (1 - 3 \cos^2 \theta + 4 \cos^4 \theta) \, .
\end{equation}
This function is plotted in Fig.~\ref{fig:angular_correlation}. It is easy to see that the angles $0$ and $\pi$ have the highest probability. 

Which types of background are to be expected? In the chart of nuclides, double beta emitters are surrounded by unstable isotopes.
Thus, the intermediate nucleus in the double beta system of interest---which also might be produced by $(p,n)$ reactions on the double beta emitter---is unstable and its beta decay into excited states will lead
to the same gamma-signature. However, the energy spectrum of the single beta decay will be continuous but overlapping with the double beta electron signal. The fraction of the beta electrons in the peak range depends on 
the energy difference of the ground states of the double beta emitter and the intermediate nucleus.
If it is small, only electrons close to the endpoint of the beta spectrum will contribute. If it is large, more electrons will contribute. A detailed estimate depends also on the quantum numbers of the ground
state of the intermediate nucleus, as allowed or forbidden beta decays will lead to different electron
energy spectra. Thus, it is essential to measure the electron energy accurately or to build a detector which is able to discriminate 
one and two electrons, typically done in detectors with tracking capabilities. 

Let us comment on the possible origin of the aforementioned $(p,n)$ reactions. Normally, there are no free protons in an underground experiment, so these reactions are not an issue. But as our consistency test relies on a very low background, even tiny contributions have to be considered. Such a contribution is the production of protons by $(n,p)$ reactions on a nucleus. Underground, high energy neutrons are dominantly produced by muon interactions in or close to the experiment. These neutrons in principle have enough energy to do $(n,p)$ reactions. A detailed estimation, however, depends on the actual cross section for $(n,p)$ reactions (mb region for neutron energies below 100~MeV) and the following $(p,n)$ reactions (mb--b region, depending on the proton energy). 

Also, external backgrounds may be an issue. The signal of a decay into excited states, however, will be a triple coincidence with well-defined energies of all involved particles. Additionally, angular correlations exist (at least between the gammas), and the total sum of particle energies must correspond to the $Q$~value of the double beta decay. These constraints make the signal search more or less background free, of course depending a little bit on the detector technology used. Especially in the case of Ge detectors with their superb energy resolution, the triple coincidence is so sharply defined that it cannot be mimicked by any other process. Decay sequences with the same gamma energies are very unlikely, furthermore the electron in the sequence will be continuous, and only a small fraction will have the right energy. Also, triple Compton events (note that we have three different energy depositions) are very unlikely. Moreover, applying the equation of Compton scattering to the three energy depositions will immediately tell whether this is consistent with Compton scattering or not.

The major background will be the $2\nu\beta\beta$ decay into the $0^+_1$ excited state, observed so far for two isotopes. As the high energy tail of the continuous two electron spectrum can mimic the signal, energy resolution becomes the crucial experimental quantity for rejecting this background.

To be more specific, the two most promising nuclides are discussed in more detail.
First, consider the case of $^{76}$Ge which is used in the GERDA and MAJORANA experiments. In addition to the signal in the sum energy 
spectrum of the electrons in form of a peak at 2039 keV as an indication of the ground state transition, the decay into the first excited $0^+$ state 
would be indicated by a sum energy of the electrons of 917 keV associated with 
two gammas of 559.1 keV and 563.2 keV, respectively. Clearly, a coincidence measurement would be preferential for this form of decay. Even ignoring the angular
correlation among the photons, this channel should be observable more or less free of background. In this case, one or two events would indicate an observation, which, however, implies 264--528 neutrinoless ground state transitions (see rates in Table~\ref{tab:exstate-alliso}). 
First Monte Carlo simulations show that for $^{76}$Ge in the case of using detectors in form of disks of 15 cm diameter and 1 cm thickness, about $60 \, \%$ of 
the gammas are expected to leave the crystal without interaction, future simulations might optimize this number. The photons might 
be detected in neighboring Ge detectors or an active medium surrounding the crystals. 

The most promising candidate is $^{150}$Nd, which is considered to be used in SNO+ and DCBA. Moreover, it is an option for SuperNEMO. 
It has a $Q$~value of 3371 keV with two gammas of 334 keV and 406 keV, respectively. As before,
the independent measurement of all these energies requires high granularity detectors, relatively small crystals in a liquid with a fair spatial resolution
or tracking devices. Again, in case of 1--2 excited state events, this would imply 120--240 events in the neutrinoless ground state 
transition. In general, depending on the actual value of the effective Majorana mass, this can be a challenging measurement.

Hence, how realistic is such an approach? Can the necessary number of counts (several hundred, see above) be reached for the decay into the ground state in future detectors?

In case the half-life $T_{1/2}$ of the isotope under consideration is much longer than the measuring time $t$, we may write the number of double beta decays as
\begin{equation}
 N_{\beta\beta} = \frac{\ln 2 a M t N_A}{T_{1/2}}\, ,
\end{equation}
where $a$ is the isotopical abundance of the nuclide of interest, $M$ is the used mass and $N_A$ is the Avogadro constant. However, in experiments, we may be confronted with background, such that there are two different possible dependencies of the expected half-life sensitivity:
\begin{equation}
 (T_{1/2})^{-1} \propto a M \epsilon t \text{~~(background free)}
\end{equation}
or 
\begin{equation}
 (T_{1/2})^{-1} \propto a \epsilon \sqrt{\frac{Mt}{B\Delta E}} \text{~~(background limited)}\, .
\end{equation}
Here, $\epsilon$ is the efficiency for detection, $B$ is the background index [typically quoted in counts/(keV kg y)], and $\Delta E$ is the energy resolution at the peak position. See~\cite{Zuber:2010bc} for a more detailed discussion on $0\nu\beta\beta$ experiments.

Consider two types of next generation detectors: a $^{76}$Ge detector, and a $^{150}$Nd detector. Two scenarios are thinkable: the Klapdor claim $T^{0\nu}_{1/2} = 2.23\times 10^{25}\,\mathrm{y}$~\cite{KlapdorKleingrothaus:2006ff} is right, so one only would have to reach this half-life. However, it is not improbable that the effective Majorana neutrino mass is as low as $50\, \mathrm{meV}$, where the inverted hierarchy begins (cf.~Fig.~\ref{fig:effective_mass}). 
Results for the running times to be accumulated in both scenarios to be able to use the proposed consistency test are given in Table~\ref{tab:running}. All parameters used are given in Table~\ref{tab:parameters}.

\renewcommand{\arraystretch}{1.5}

\begin{table}
 \centering
\caption{Running times to be accumulated for the two types of next-generation detectors discussed in this paper, to be able to use the proposed consistency test. The two thinkable scenarios are described in the text.}\label{tab:running}

\begin{tabular}{lcc}\hline
 &$^{76}$Ge  & $^{150}$Nd  \\ \hline\hline
Klapdor's claim $T^{0\nu}_{1/2} = 2.23\times 10^{25}\,\mathrm{y}$~\cite{KlapdorKleingrothaus:2006ff}: & & \\
\hphantom{xxx} background free & $ Mt = 2.0\, \mathrm{ton}\,\mathrm{y}$ & $Mt = 3.4  \, \mathrm{ton}\,\mathrm{y}$\\
\hphantom{xxx} background limited & $Mt = 121 \, \mathrm{ton}\,\mathrm{y}$ & $Mt = 30.9  \, \mathrm{kton}\,\mathrm{y}$\\ \hline
$\ema = 50\,\mathrm{meV}$: & & \\
\hphantom{xxx} background free & $Mt = 48.3  \, \mathrm{ton}\,\mathrm{y}$& $Mt = 14.2  \, \mathrm{ton}\,\mathrm{y}$\\
\hphantom{xxx} background limited & $Mt = 69.9  \, \mathrm{kton}\,\mathrm{y}$ & $ Mt = 543  \, \mathrm{kton}\,\mathrm{y}$\\ \hline
\end{tabular}
\end{table}

\renewcommand{\arraystretch}{1.5}

\begin{table}
 \centering
\caption{Parameters used in the calculations. The values for $^{76}$Ge are as currently taken for the GERDA experiment (phase I)~\cite{Smolnikov:2010zza}, the values for $^{150}$Nd are as expected for SuperNEMO~\cite{Ohsumi:2008zz}. Nuclear matrix elements are IBM-2~\cite{Barea:2009zza,iac2010,iac2011}, and the phase space factors are taken from~\cite{Doi:1985dx}.}\label{tab:parameters}

\begin{tabular}{lcc}\hline
 &$^{76}$Ge  & $^{150}$Nd  \\ \hline\hline
Isotopical abundance $a\, [m_u^{-1}]$ & $85 \%$ \cite{Smolnikov:2010zza}& $90 \%$\\
Efficiency $\epsilon$ & $ 0.60$ & $0.30$~\cite{Ohsumi:2008zz}\\
Number of decays to ground state $N_{\beta\beta}$ & $264$ & $120$ \\
Background $B \, [\mathrm{counts/(keV\, kg\, y)}]$ & $0.01$ \cite{Smolnikov:2010zza} & $0.02$~\cite{Ohsumi:2008zz}\\
Energy resolution at peak position $\Delta E$ & $3 \, \mathrm{keV}$ \cite{Smolnikov:2010zza} & $ 4 \, \%$~\cite{Ohsumi:2008zz} \\
$\mathcal{M}^{0\nu} $ & $5.465$  &  $2.321$ \\
$G^{0\nu} \, [\mathrm{y}^{-1} \, \mathrm{eV}^{-2}]$ & $0.25 \times 10^{-25}$ & $8.03 \times 10^{-25}$ \\ 
typical mass $[\mathrm{kg}]$& $1000$ & $200$ \\ \hline
\end{tabular}

\end{table}

How big may the half-life maximally be, so that the proposed consistency check can be used in a running time of, say, 10 years? This means that we need at least 264 or 120 decays to the ground state in $^{76}$Ge or $^{150}$Nd, respectively. Assuming a background-free experiment, and using the values from Table~\ref{tab:parameters}, for a $^{76}$Ge detector of $1\, \mathrm{ton}$, the maximal half-life where this number of decays to the ground state is reached is
\begin{equation}
 T_{1/2} = 1.11\times 10^{26} \, \mathrm{y}\, .
\end{equation}
This corresponds (using the matrix elements and phase space factors in Table~\ref{tab:parameters}) to an effective Majorana neutrino mass of $0.11 \, \mathrm{eV}$. For a $^{150}$Nd detector of $200 \, \mathrm{kg}$, the maximal half-life is
\begin{equation}
 T_{1/2} = 1.31\times 10^{25} \, \mathrm{y}\, ,
\end{equation}
which corresponds (using again Table~\ref{tab:parameters}) to an effective Majorana neutrino mass of $0.14 \, \mathrm{eV}$. 
Thus, for detectors of the size of several hundred kilograms up to $1\,\mathrm{ton}$, the method should work down to about $100\, \mathrm{meV}$.

\section{Summary and conclusions}

\begin{figure}
 \centering
\includegraphics[width=0.76\textwidth]{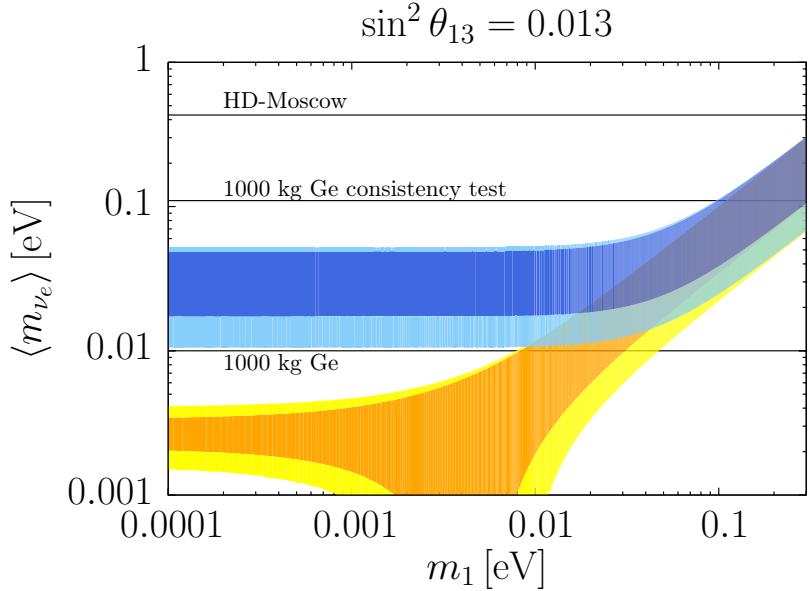}
 \caption{The effective Majorana mass \ema\ as a function of the lightest neutrino mass eigenvalue for inverted (blue/light blue) and normal (orange/yellow) hierarchy. In the quasi degenerate case, both hierarchies overlap. Bold colors denote the best fit values range (varying the $CP$ phases), light colors give the corresponding $3 \sigma$ values range. The best fit value $\ema = 0.34\, \mathrm{eV}$ obtained in the Heidelberg-Moscow experiment~\cite{KlapdorKleingrothaus:2004wj} is marked. A future $1 \, \mathrm{ton}$ $^{76}$Ge experiment~\cite{Schonert:2005zn} could probe the inverted hierarchy down to $\ema = 0.01\, \mathrm{eV}$. Using this experiment for the consistency test, the probed value for \ema\ would be higher, due to the lower rate of the decay to excited states.}\label{fig:effective_mass}
\end{figure}

We have proposed a method to check within a single experiment whether a possibly observed signal in a future $0\nu\beta\beta$ detector is really due to $0\nu\beta\beta$ or due to some unknown nuclear line. This question will arise if a positive signal is seen in a second-generation detector like GERDA, which is about to start data-taking and which will then reach and exceed the sensitivity of the Heidelberg-Moscow experiment to a certain point. Usually, it is argued that another second generation experiment with a different isotope can settle the question. We pointed out that it is also possible to combine effort into one large detector, instead of using various different isotopes. Such a detector would therefore serve different purposes: It would have sensitivity to lower \ema\ and it would be able to check a claim for higher \ema\ due to the very characteristic features described in this paper. See Fig.~\ref{fig:effective_mass} for the reach for both modes. It is clear from the figure that the proposed consistency test is only viable for relatively large effective neutrino mass. Should we really have to cover the whole inverted hierarchy, or even go down to normal hierarchy, one large detector may not be viable due to the large amount of material needed, and a set of various isotopes in smaller detectors may be preferable to check for consistency.
We mentioned also that our proposed method may allow a better extraction of the effective neutrino mass, since common errors in the calculations of nuclear matrix elements may cancel in ratios. Of course, a deeper discussion of the consistency test is necessary before a particular experimental setup may be proposed. However, our considerations should have proven the viability of such a method. 

\section*{Acknowledgements}
This work has been supported by the DFG Sonderforschungsbereich Transregio 27 ``Neutrinos and beyond -- Weakly interacting particles in Physics, Astrophysics and Cosmology''. MD is supported by the International Max Planck Research School (IMPRS) ``Precision Tests of Fundamental Symmetries'' of the Max Planck Society. MD and ML thank M.\ Hirsch for useful and inspiring discussions. We thank B. Lehnert for the help with the simulations.

\clearpage

\begin{small}
\bibliographystyle{./utcaps_mod}
\bibliography{./BB-SingleIso.bib}
\end{small}

\end{document}